\documentclass[options]{JHEP3}
\usepackage{amsmath,amssymb}
\title{Propagators and WKB-exactness in the plane wave limit
of $AdS\times S$}
\author{Danilo E. Diaz, Harald Dorn
\\ Humboldt-Universit\"at zu Berlin, Institut f\"ur Physik
\\Newtonstr.15, D-12489 Berlin
\\E-mail: \email{ddiaz@physik.hu-berlin.de, dorn@physik.hu-berlin.de}}

\abstract{Green functions for the scalar, spinor and vector fields
in a plane wave geometry arising as a Penrose limit of $AdS\times
S$  are obtained. The Schwinger-DeWitt technique directly gives
the results in the plane wave background, which turns out to be
WKB-exact. Therefore the structural similarity with flat space
results is unveiled. In addition, based on the local character of
the Penrose limit, it is claimed that for getting the correct
propagators in the limit one can rely on the first terms of the
direct geodesic contribution in the Schwinger-DeWitt expansion of
the original propagators . This is explicitly shown for the
Einstein Static Universe, which has the same Penrose limit as
$AdS\times S$ with equal radii, and for a number of other
illustrative cases.}  \preprint{HU Berlin-EP-04/34}

\begin{document}

\section{Introduction}
Recently, the study of strings in plane wave backgrounds has
received a lot of attention. These activities are due to the
observation \cite{BMN02} that, via a special limit of the standard
AdS/CFT correspondence, string theory on a certain plane wave
background corresponds to a large R-charge subsector of the ${\cal
N}=4$ super Yang-Mills gauge field theory. In contrast to the
original $AdS_5\times S^5$, in these plane wave backgrounds the
exact quantization of strings is known. This allows tests of the
correspondence including genuine stringy properties.

Related to these considerations also the field theoretical
properties of plane wave backgrounds became relevant. In
particular the propagators, both bulk to bulk and bulk to
boundary, for $AdS_5\times S^5$ and for the plane wave arising in
a Penrose limit should play a crucial role in understanding the
degeneration of the holographic picture from a 4- to a
1-dimensional boundary. In spite of several attempts \cite{KiPi02,
CB,DS03}, this issue is up to now not completely clarified.

The scalar propagator in the relevant plane wave has been
constructed for generic mass values via direct mode summation in
\cite{MSS02}. In addition, there have been observed structural
similarities with the flat space propagator and their possible
role in guessing the higher spin propagators was stressed. The
alternative route via the limiting behavior of the
$AdS_5\times S^5$ propagator was taken in \cite{DSS03} for the
conformally coupled scalar.\\

In the present paper we want to address the propagators in the
plane wave background along the line of the Schwinger-DeWitt
construction. This technique, based on an expansion near the light
cone, has a long history. It has been successfully applied to the
propagator construction in various specific backgrounds as well as
to issues related to near light cone and anomaly problems in
generic backgrounds. It lies in the heart of most regularization
techniques of QFT in curved spaces (see, e.g.~\cite{BD84}). Our aim
here is to explain the above mentioned structural similarities to
the flat space case by the termination of the underlying WKB
expansion and to make progress in the explicit construction for
higher spin cases. We will also explore the alternative approach
to derive the plane wave propagators as a limiting case of
propagators in spaces which in a Penrose limit yield the plane
wave. For this we relate our results to information on propagators
in Einstein Static Universe ($ESU$) available in the literature.\\

The paper is organized as follows. After collecting some
preliminaries on plane waves, $AdS$ and $ESU$ in section 2, we
apply in section 3 the Schwinger-DeWitt technique to construct the
plane wave propagators for the scalar, spin 1/2 and the spin 1
gauge field propagator. Section 4 is devoted to some comments on
known results on propagators and Schwinger-DeWitt kernels in $ESU$
and their relation via a Penrose limit to the spin 0 and spin 1/2
results of the previous section. Section 5 reproduces for the
conformal flat cases of $AdS_{p+1}\times S^{q+1}$ and Weyl
invariant coupling of scalars the results of \cite{DSS03} within
the Schwinger-DeWitt technique and comments on the role of direct
and indirect geodesic contributions before and after taking the
Penrose limit to the plane wave. In addition we find a relation
between the propagators in a special conformal non-flat situation 
and a flat one via a contour integral. This special non-flat
situation concerns just the special values for curvature radii and
mass values which still allowed an explicit summation in
\cite{DSS03}. We end with a summary and some conclusions. Various
technical details are collected in a set of appendices.

\section{Penrose limit: plane wave background}

The particular plane wave background to be considered is the conformally
flat one obtained as a Penrose limit of $AdS_5\times
S^5$~\cite{BFHP02,BMN02} with equal radii, although at some stages
the results can be adapted to other dimensions by just varying the
number of transverse directions $\vec{x}$. The line element is
given by
\begin{equation}
\label{pp_metric} ds^2 = 2 du dv - \vec{x}^2 du^2+ d\vec{x}^2.
\end{equation}

As noticed by Penrose~\cite{Pen76}, this limit is nothing but an
adaptation to pseudo-Riemannian manifolds of the standard
procedure of taking tangent space limit, the main difference being
that when applied to a null geodesic it results in curved space,
namely a plane wave. One could as well end up with flat space,
but the generic situation is a plane wave. It is this zooming into
the neighborhood of the null geodesic what gives the Penrose limit
a local character.

Recently, Penrose limits of a whole variety of space-times has
been thoroughly studied (see, e.g. \cite{KiPi02} and reference[13]
therein). The particular plane wave metric (\ref{pp_metric})
together with a RR-flux corresponds to a maximally supersymmetric
solution of Type II-B SUGRA, as first found by BFHP~\cite{BFHP02}.
This very Type II-B SUGRA background can also be obtained as a
Penrose limit of the less supersymmetric $AdS_5\times
T^{1,1}$~\cite{AdSxT}, and surely from many other backgrounds.
Now, as far as one is interested only in the metric, the spacetime
with the same Penrose limit, which ought to be considered the
conceptually simplest one, is the Einstein Static Universe
$ESU_{10}$. In parts of the following discussion we will benefit
from this fact.
\subsection{Anti-deSitter$\times$Sphere}

Let us start with $AdS_{p+1}$ in global coordinates and with the
($q+1$)-sphere parametrized in terms of a ($q-1$)-sphere ($a$ is
the common radius of $AdS$ and the sphere)
\begin{equation}
\label{global} ds_{AdS_{p+1}\times S^{q+1}}^2 =
a^2\left(-dt^2\cosh^2\rho + d\rho^2+\sinh^2\rho d\Omega_{p-1}^2
+\cos^2\theta d\psi^2 + d\theta^2 +\sin^2\theta
d\Omega_{q-1}^{'2}\right).
\end{equation}
Now one focuses on the immediate neighborhood of a null geodesic that
remains at the center of $AdS_{p+1}$ while it wraps an equator of
$S^{q+1}$, say $t=\psi=u$ (affine parameter along the null ray) and
$\rho=\theta=0$. Introducing local coordinates
\begin{equation}\label{loccoor}
t=u\qquad\qquad\psi=u+\frac{v}{a^2}\qquad\qquad\rho=\frac{x}{a}
\qquad\qquad\theta=\frac{y}{a}
\end{equation}
and expanding in inverse powers of the radius, one gets
\begin{equation}
\label{local}
ds_{AdS_{p+1}\times S^{q+1}}^2 = 2 du dv - ( x^2 +
y^2 ) du^2+ dx^2 + x^2 d\Omega_{p-1}^2 + dy^2 +
y^2d\Omega_{q-1}^{'2} + O(a^{-2}),
\end{equation}
so that in the limit $a\rightarrow\infty$ , blowing up the
neighborhood and collecting the flat transverse directions into
$\vec{x}$, one ends up with the plane wave metric (\ref{pp_metric}).
\subsection{Einstein Static Universe}

To see that the same plane wave results from $ESU_n$, topologically
$R\times S^{n-1}$, let us conveniently parametrize the ($n-1$)-sphere in
terms of a ($n-3$)-sphere
\begin{subequations}
\begin{eqnarray}
ds_{ESU_n}^2 &=& a^2(-dt^2+ d \Omega_{n-1}^2)\\
&=& a^2\left(-dt^2+d\alpha^2+\cos^2\alpha \; d\beta^2
+\sin^2\alpha \;d \Omega_{n-3}^2\right).
\end{eqnarray}
\end{subequations}

This time the null geodesic will be the one given by $t=\beta=u$
(affine parameter along the null ray) and $\alpha=0$, and the
local coordinates in its neighborhood
\begin{equation}t=u\qquad\qquad\beta=
u+\frac{v}{a^2}\qquad\qquad\alpha=\frac{r}{a}\,.\end{equation}
Then, expanding the line element
\begin{equation}
ds^2_{ESU_n} = 2 du dv - r^2 du^2+ dr^2 + r^2 d\Omega_{n-3}^2 +
O(a^{-2})\end{equation} and letting $a \rightarrow \infty$ one
gets again the plane wave metric (\ref{pp_metric}).

That both Penrose limits give the same metric can be easily
understood if one remembers that there is a conformal map that
allows for a Penrose diagram for $AdS_{p+1}\times S^{q+1}$.
Defining $\tan{\vartheta}\equiv\sinh{\rho}$ in (\ref{global}), one
obtains that both metrics are related by\footnote{Obviously, there
is an obstruction
  to this argument if the null geodesic, on which one focuses in the Penrose
limit, reaches the boundary of $AdS\times S$ where the conformal factor
becomes singular. It is precisely in this situation when the null geodesic is
totally contained in $AdS$ and the Penrose limit of $AdS\times S$ gives
just Minkowski space.}
\begin{equation}
ds_{AdS_{p+1}\times S^{q+1}}^2 = \frac{1}{\cos^2{\vartheta}}
ds_{ESU_{p+q+2}}^2.
\end{equation}

Now, in the local coordinates (\ref{loccoor}) near the null geodesic at the
center of $Ad_{p+1}$ we have $\vartheta=\frac{x}{a}+O(a^{-3})$ and the
conformal factor $\cos^{\pm2}\vartheta=1 + O(a^{-2})$, therefore up to
$O(a^{-2})$ both metric are equivalent, i.e. the RHS of
(\ref{local}) holds for both backgrounds. Consequently, in the
limit $a \rightarrow \infty$ the resulting metrics coincide. Notice that this
time we had a different parametrization of the $(n-1)-$sphere in $ESU_n$ and
different local coordinates, but their departures are scaled away in the
plane wave limit resulting in the same spacetime. This is again a
manifestation of the inherent locality of the Penrose limit.

\section{Propagators in the plane wave}

The Feynman scalar propagator in the plane wave background has already been
obtained by explicit summation of the eigenmodes in recent
works~\cite{MSS02,KiPi02}.
Here we will treat it differently using the
Schwinger-DeWitt technique which admits a readily
generalization to the spinor and vector fields.
\subsection{Scalar propagator: Schwinger-DeWitt proper-time }

The scalar Feynman propagator in the curved background of the plane wave is
the solution of the wave equation with a point-like source
\begin{equation}
\left(\square - m^2\right)G(x,x')= \delta(x,x')
\end{equation}
together with appropriate boundary  conditions. Here $\delta(x,x')$ denotes
the invariant $\delta$-function. The Schwinger-DeWitt
proper-time representation for the Feynman propagator~\cite{DeW65}, which
incorporates the
Feynman boundary conditions by the $i0^+$ prescription, is based on the
formal solution
\begin{equation}\frac{1}{\square - m^2+i0^+}=-i \int_0^{\infty} ds\,
e^{-is\,m^2-s0^+} \, e^{is\,\square}\,.
\end{equation}
The Schwinger-DeWitt kernel (the kernel of the exponentiated
operator),
\begin{equation}
K(x\,,x'\mid s)\equiv\langle x \mid e^{is\,\square} \mid x'
\rangle=e^{is\,\square}\delta(x,x')
\end{equation} satisfies the following ``Schr\"odinger equation'' and initial
condition
\begin{subequations}
\begin{eqnarray} \left( i\partial_s+ \square \right) K(x,x'\mid s)=0\\
 K(x,x'\mid 0)=\delta(x,x').
\end{eqnarray}
\end{subequations}
A WKB-inspired ansatz for the solution, meant to be only an
asymptotic one, is
\begin{equation}K(x,x'\mid s)=\frac{i}{(4\pi is)^{\frac{d}{2}}}\,
\triangle^{\frac{1}{2}} e^{\,i\sigma/2s}\, \Omega(x,x'\mid s)
+...
\end{equation}
where\,$\sigma(x,x')$ is the geodetic interval (one half the
geodesic distance squared between the two points),
\begin{equation}
\triangle(x,x')[g(x) g(x')]^\frac{1}{2}\equiv-\mbox{det}
(-\frac{\partial^2\sigma}{\partial x^{\mu}\partial x^{'\nu}})
\end{equation}
is the Van Vleck-Morette determinant( an important improvement of
the WKB ansatz). $\Omega(x,x'\mid s)$ has a power expansion in the
 proper time $s$
\begin{equation}\Omega(x,x'\mid s)=\, \sum_{n=0}^{\infty} (is)^n\,
a_n(x,x'),
\end{equation}
whose coefficients $a_n(x,x')$ are regular functions in the
coincidence limit $x\rightarrow x'$, and finally the ellipsis
stands for indirect geodesic contributions. The coefficients,
sometimes referred to as HaMiDeW coefficients, must satisfy the
recursion relation
\begin{equation}
\label{rec}
(n+1)\,a_{n+1}+\partial^\mu\sigma\;\partial_\mu
a_{n+1}=\triangle^{-\frac{1}{2}}\,\square
\,(\triangle^{\frac{1}{2}}a_n)
\end{equation}
starting with $\partial^\mu\sigma\;\partial_\mu a_0=0$ and
$a_0(x,x)=1$. For the present scalar case, the chain of HaMiDeW
coefficients trivially starts with $a_0(x,x')=1$.

Now we are in position to apply this construction to the plane wave
background. From the geodetic interval between two generic points
(see appendix~\ref{appendixC}) one obtains the Van Vleck-Morette determinant.
The important ingredients are
\begin{subequations} \label{pp_ing}
\begin{eqnarray}
g(x) &=& -1\\
\sigma(x,x') &=& (u-u')\left[v-v'+\frac{\vec{x}^2+
\vec{x}'^2}{2}
\,\cot{(u-u')}-\vec{x}\cdot\vec{x}'\,\csc{(u-u')}\right]\\
\triangle(x,x') &=& \left[\frac{u-u'}{\sin{(u-u')}}\right]^{d-2}.
\end{eqnarray}
\end{subequations}
With this at hand, one can check that
$\triangle^{\frac{1}{2}}(x,\cdot)$ is harmonic, i.e. $\square
\,\triangle^{\frac{1}{2}}(x,\cdot)=0$, because
$\triangle(x,\cdot)$ is a function only of $u$ and the inverse
metric has $g^{uu}=0$, so that the recurrence relations are
satisfied by $a_n(x,x')=\delta_{0,n}$ . Thus the only non-zero
coefficient in the expansion is just the first one. That is why we
say that the scalar Schwinger-DeWitt kernel in the plane wave
background is leading-WKB exact. The kernel and the Green
function, after performing the proper time integral, are then
given by\footnote{ As usually, the Feynman Green function should
be understood as the boundary value of a function which is
analytic in the upper-half $\sigma$ plane, so that in fact $\sigma
+ i0^+$ is meant in what follows.}
\begin{subequations}\label{pp_results}
\begin{eqnarray}
K(x,x'\mid s)&=&\frac{i\triangle^{\frac{1}{2}}}{(4\pi
is)^{\frac{d}{2}}}\;e^{\,i\sigma/2s}\\
G(x,x') &=&\frac{-i\pi\triangle^{\frac{1}{2}}}{(4\pi
i)^{\frac{d}{2}}}\;\left(\frac{2 m^2}{\sigma}\right)^{\frac{d-2}{4}}\;
H^{(2)}_{\frac{d}{2}-1}\left(\left[-2m^2\sigma\right]^{\frac{1}{2}}\right).
\end{eqnarray}
\end{subequations}

One can get Minkowski space by rescaling $u\rightarrow \mu u,
v\rightarrow v/\mu $ and letting $\mu$ go to zero. The effect of
this in (\ref{pp_ing}, \ref{pp_results}) is $\triangle\rightarrow 1$ and
$2\sigma\rightarrow 2(u-u')(v-v') + (\vec{x}-\vec{x}')^2$ and
\begin{subequations}
\begin{eqnarray}
K_{M}(x,x'\mid s)&=&\frac{i}{(4\pi
is)^{\frac{d}{2}}}\;e^{\,i\sigma/2s}\\
G_{M}(x,x') &=&\frac{-i\pi}{(4\pi
i)^{\frac{d}{2}}}\;\left(\frac{2 m^2}{\sigma}\right)^{\frac{d-2}{4}}\;
H^{(2)}_{\frac{d}{2}-1}\left(\left[-2m^2\sigma\right]^{\frac{1}{2}}\right).
\end{eqnarray}
\end{subequations}

The difference between the two results, apart from the fact that
the geodetic interval is of course different, is that for the
plane wave we get a nontrivial Van Vleck-Morette determinant. The
analogy with the Minkowski case observed in~\cite{MSS02} is thus
fully explained by the leading-WKB exactness of the plane wave
background. The coincidence limit of our results, where the
coefficients become local functions of curvature
invariants~\cite{DeW65,Chr76}, is consistent with the fact that
for the plane wave background there are no non-vanishing curvature
invariants~\cite{HS90}.

Finally, for the massless scalar one can take the massless limit
in both expressions to get\footnote{
The geometrical meaning of the quantity $\Phi$ is explained in
appendix~\ref{appendixA}.}
\begin{equation}
\label{D}
D(x,x') = \frac{-i\,\Gamma(d/2-1)}{2(2\pi)^{d/2}}
\triangle^{\frac{1}{2}}
\left(\frac{1}{\sigma}\right)^{\frac{d-2}{2}}=  \frac{-i\,\Gamma(d/2-1)}
{2(2\pi)^{d/2}}\left(\frac{1}{\Phi}\right)^{\frac{d-2}{2}}
\end{equation}
\begin{equation}
D_M(x,x') = \frac{-i\,\Gamma(d/2-1)}{2(2\pi)^{d/2}}\left(
\frac{1}{\sigma}\right)^{\frac{d-2}{2}}.
\end{equation}

\subsection{Spinor field: leading-WKB-exactness}

One might guess that the similarity with Minkowski space kernel and  Green
function still holds for higher spin fields. Now we will turn our attention to
the spin $\frac{1}{2}$ case.

The spinor Green function is now a bi-spinor which satisfies the
Dirac equation with a point-like source
 \begin{equation}
\left[ \gamma^\mu(x) \nabla_\mu + m \right]\mathbb{S}(x,x')=
\delta(x,x')\mathbb{I},
\end{equation}
where $\gamma^\mu(x)$ are the curved space Dirac matrices and
$\nabla_\mu$ is the spinor covariant derivative (see appendix
\ref{appendixB}).

To apply the Schwinger-DeWitt technique one introduces an
auxiliary bi-spinor $\mathbb{G}(x,x')$ defined by
 \begin{equation}
\mathbb{S}(x,x') = \left(\gamma^\mu(x) \nabla_\mu - m\right)\mathbb{G}(x,x')
\end{equation}
to obtain the following wave equation for $\mathbb{G}(x,x')$
\begin{equation}
(\square-\frac{R}{4} - m^2)\mathbb{G}(x,x')=
\delta(x,x')\mathbb{I},
\end{equation}
where $R$ is the scalar curvature.

Now one can apply the Schwinger-DeWitt construction as in the
scalar case, but this time the auxiliary Green function
$\mathbb{G}(x,x')$, the kernel $\mathbb{K}(x,x'\mid s)$ as well as
the HaMiDeW coefficients $\mathbb{A}_n(x,x')$ are bi-spinors and
the recurrence relations (\ref{rec}) involve the spinor covariant derivative.
One starts with $\mathbb{A}_0(x,x')=\mathbb{U}(x,x')$, the spinor parallel
transporter along the geodesic connecting the two points
(appendix \ref{appendixB}).

For the plane wave background (\ref{pp_metric}) one can check that
again the recurrence relations are satisfied by
$\mathbb{A}_n(x,x')=\delta_{n,0}\mathbb{U}(x,x')$, the reason
being that $\triangle^{\frac{1}{2}}(x,\cdot)\mathbb{U}(x,\cdot)$
is harmonic, with respect to the spinor D'Alembertian.

Therefore, the spinor kernel and the spinor auxiliary Green
function are leading-WKB exact and can be written in terms of the
respective scalar quantities
\begin{subequations}
\begin{eqnarray}
\mathbb{K}(x,x'\mid s) &=& K(x,x'\mid
s)\,\mathbb{U}(x,x'),\\
\mathbb{G}(x,x') &=& G(x,x')\,\mathbb{U}(x,x').
\end{eqnarray}
\end{subequations}

Flat space results can also be recovered as in the scalar
case, taking into account that in the limit
$\mathbb{U}(x,x')\rightarrow\mathbb{I}$. The similarity with flat
space result is still present, the only additional nontrivial
piece being the spinor geodesic parallel transporter, and is better appreciated
in terms of the kernel and the auxiliary Green function, so we do not show
the explicit expression for $\mathbb{S}$.

\subsection{Vector field: next-to-leading-WKB-exactness}

Let us examine the Maxwell field. Now we have additional
complications due to the gauge freedom, so we add a gauge fixing
term $-\frac{1}{2\xi}(\nabla_\mu A^\mu)^2$ in the action
to get an invertible differential operator
\begin{equation}
\left[g_{\mu\rho}\square-R_{\mu\rho}-(1-\xi^{-1})\nabla_\mu
\nabla_\rho\right] G^{\rho}_{\nu'}(x,x')= \delta(x,x')g_{\mu\nu'}(x),
\end{equation}
where the Ricci tensor $R_{\mu\nu}$ arises from the commutator of
the covariant derivatives. Its only non-vanishing component in the plane wave
geometry is $R_{uu}=d-2$. This can be easily obtained from the Christoffel
symbols (\ref{Christ}).

In the Feynman gauge $\xi=1$, corresponding to a ``minimal'' wave
operator in the sense of Barvinsky and Vilkovisky ~\cite{BV85},
one can work out a Schwinger-DeWitt construction and this time we
have to deal with bi-vectors. The recurrences are slightly changed to
\begin{equation}
(n+1)\,a_{n+1\,\mu\nu'}+\partial^\rho\sigma\;\nabla_\rho
a_{n+1\,\mu\nu'}=\triangle^{-\frac{1}{2}}\,\square
\,(\triangle^{\frac{1}{2}}a_{n\,\mu\nu'})-R_{\mu}^{\;\rho}a_{n\,\rho\nu'},
\end{equation}
and the chain of HaMiDeW coefficients starts with the vector
geodesic parallel transporter
$a_{0\,\mu\nu'}(x,x')=P_{\mu\nu'}(x,x')$ (appendix \ref{appendixC}) .

What one can show in this case is that the recurrences are solved
by
\begin{equation}
a_{n\,\mu\nu'}(x,x')=\left\{\begin{array}{ll}
P_{\mu\nu'}(x,x'), & n=0,\\
(2-d)\delta_{u\mu}\delta_{u'\nu'}\frac{\tan\frac{u-u'}{2}}{\frac{u-u'}{2}},
& n=1,\\
0, & n \geq 2.\end{array}\right.
\end{equation}

 The vector kernel is then
\begin{equation}
\label{vecK}
K^{\mu}_{\;\nu'}(x,x'\mid s)=\frac{i\triangle^{\frac{1}{2}}}{(4\pi
is)^{\frac{d}{2}}}\;e^{\,i\sigma/2s}\left(\delta^{\mu}_{\;\rho}-is
\frac{\tan\frac{u-u'}{2}}{\frac{u-u'}{2}}R^{\mu}_{\;\rho} \right)
P^{\rho}_{\;\nu'}(x,x'),
\end{equation}
and the Green function can be written in terms of the massless
scalar Green function as
\begin{equation}
G^{\mu}_{\;\nu'}(x,x')=\left(D(x,x')\,
\delta^{\mu}_{\;\rho}-\frac{1}{4\pi\cos^2\frac{u-u'}{2}}
 Q(x,x') R^{\mu}_{\;\rho}\right)P^{\rho}_{\;\nu'}(x,x'),
\end{equation}
where the functional dependence of $Q$ on $u-u'$ and $\sigma$ is precisely the
same as in $D$ but in two dimensions less (see \ref{D}), i.e.
\begin{eqnarray}
Q(x,x') &=& \frac{-i\,\Gamma(d/2-2)}{2(2\pi)^{d/2-1}}
 \left[\frac{u-u'}{\sin{(u-u')}}\right]^{\frac{d-4}{2}}
\left(\frac{1}{\sigma}\right)^{\frac{d-4}{2}}
= \frac{-i\,\Gamma(d/2-2)}
{2(2\pi)^{d/2-1}}\left(\frac{1}{\Phi}\right)^{\frac{d-4}{2}}
\end{eqnarray}
That we should not expect leading-WKB exactness this time can be
seen by examining the coincidence limit $x\rightarrow x'$, where
general results~\cite{DeW65,Chr76} are available. In particular for the plane
wave under consideration, one must have $a_{1\,\mu\nu}(x,x)=-R_{\mu\nu}(x)$,
and this can be readily checked in (\ref{vecK}) remembering that the
coincidence limit of the vector parallel transporter is just the metric tensor,
$P_{\mu\nu}(x,x)=g_{\mu\nu}(x)$.

After all, we obtained the minimal departure: next-to-leading WKB-
exactness. This time, the similarity with flat space results is
still present although obscured by an additional term. The flat
space limit can be taken as in the preceding two cases, this time
the vector parallel transporter goes to the metric tensor and the
$a_1$ coefficient together with the Ricci tensor go to zero to end
up with the usual Minkowski space results in Feynman gauge, that
is, the metric tensor times the massless scalar propagator.

\section{Propagators in $ESU$: resummation and Penrose limit}

One can take advantage of the fact that the ESU has the same Penrose limit
and try to take the limit directly in the Green functions for ESU where
some results are available in the literature. How does the limit work directly
on the propagators is apparently not easy to see in the mode summation form.
But, after a resummation things might get clearer.
 The resummation we will explore is the one implicit in the so
called ``duality spectrum-geodesics''. That is, the kernel can be
written either as an eigenfunction expansion or as a ``sum over
classical paths''~\cite{Cam90}.

\subsection{Scalar field in $ESU$}
The resummation is implicit in the following form for the scalar
Green function in $ESU_4$ obtained by Dowker and
Critchley~\cite{DC77}(see also \cite{Cam90}) based on the
Schwinger-DeWitt technique. The Schwinger-DeWitt kernel, as well
as the heat kernel, factorizes for a product space and since
$ESU_4$ is nothing but $R\times S^3$ one just needs
 the free kernel for the time direction
$K_R(t,t'\mid s)=\frac{i}{(4\pi is)^{1/2}}\;e^{-ia^2(t-t')^2/4s}$
and the kernel for the 3-sphere. The whole problem reduces to
finding $K_{S^3}$ and one can show that the 3-sphere is
leading-WKB exact\footnote{ The odd-dimensional spheres turn out
to be WKB exact after factorizing a constant phase involving the
scalar curvature. This phase can in turn be absorbed in the
definition of the differential operator and its effect in the
Green function is just a shift in the mass. This must be taken
into account when comparing the results in \cite{DC77} with those
in \cite{Cam90}}. The only complication is that, due to the
compactness of the sphere, one has multiple geodesics in addition
to the direct one so that one has to include indirect geodesic
contributions which restore the periodicity on the sphere
\begin{equation}
K_{S^3}(q,q'\mid s)=\, \sum_{n=-\infty}^{\infty}K^0_{S^3}(\chi+2\pi n a\mid s),
\end{equation}
where $\chi$ is the length of the shortest arc connecting the two points $q,
q'$ on the 3-sphere and
\begin{equation}
K^0_{S^3}(\chi\mid s)=\frac{1}{(4\pi
is)^{\frac{3}{2}}}\,\triangle^\frac{1}{2}\;e^{\,i\chi^2/4s+\,is/a^2},
\end{equation}
with the Van Vleck-Morette determinant for the sphere resulting in
$\triangle^{\frac{1}{2}}=\frac{\chi/a}{\sin(\chi/a)}$.
The corresponding Green function for $ESU_4$ is also given by direct plus
indirect geodesic contributions
\begin{eqnarray}
G_{ESU_4}(x,x')=\, \sum_{n=-\infty}^{\infty}G^0_{ESU_4}(t-t',\chi+2\pi n a)
\\
G^0_{ESU_4}(t-t',\chi)=\frac{i\triangle^\frac{1}{2}}{8\pi}\,
\left(\frac{m^2-a^{-2}}{2\sigma}\right)^{\frac{1}{2}}\;H^{(2)}_1\left(
\left[-2(m^2-a^{-2})\sigma\right]^{\frac{1}{2}}\right),
\end{eqnarray}
 where the direct geodetic interval is $\sigma=\frac{ -a^2(t-t')^2+\chi^2}{2}$.
Now  one can take the Penrose limit (see appendix~\ref{appendixA}),  and the
result is that only the  direct geodesic contribution survives the limit to
give precisely the plane wave results (\ref{pp_results}) for $d=4$.
The indirect geodesic terms become rapidly oscillating or exponentially
decaying. Therefore they vanish as a distribution for
$a\rightarrow \infty$. This is similar to the flat space limit of $ESU_4$
discussed in~\cite{DC77}.

This construction can be generalized to higher dimensional $ESU_n$. For
odd-dimensional spheres the Schwinger-DeWitt kernel is WKB exact~\cite{Cam90}
and for even-dimensional spheres one only has an asymptotic expansion,
but in all cases the only term that survives the Penrose limit is the first
coefficient in the direct geodesic contribution and this can be seen in the
asymptotic expansion, all other terms are suppressed by inverse powers of
the radius or are rapidly oscillating.
\subsection{Spinor field in ESU}

For $ESU_4$, Altaie and Dowker~\cite{DA78} obtained the spinor S-D kernel and
the spinor Green function. To our purposes it suffices to take a look at the
spinor S-D kernel, which due to the compactness of the 3-sphere is again a sum
over all geodesics connecting the two points, with the direct term
\begin{equation}
\mathbb{K}^0_{ESU_4}(x,x'\mid s)=\frac{i}{(4\pi
is)^2}\,\triangle^\frac{1}{2}\;e^{\,i(\chi^2-a^2(t-t')^2)/4s}\,
\left(1-\frac{is\tan(\chi/a)}{a\chi}
\right)\, \mathbb{U}(x,x').
\end{equation}
In the Penrose limit (see appendix \ref{appendixA}) one gets again the same
behavior, i.e. only the first
coefficient in the direct geodesic term survive and everything else is
suppressed as in the scalar case.

One can follow this construction using the spinor kernel for the
higher-dimensional spheres, already calculated by
Camporesi~\cite{Cam92}, and one gets again agreement with
our previous results from direct computation in the plane wave background.
In all cases, the relevant information is contained in the S-D asymptotic
(S-D stands either for Schwinger-DeWitt or for short-distance), the rest is
just scaled away in the Penrose limit. This is precisely the resummation we
were looking for.

\section{Plane wave propagators via Penrose limit of $AdS\times S$}

The key tool for the previous results was the resummation implicit in
the Schwinger-DeWitt asymptotics.
So, this could be the recipe to obtain the limiting values of the Green
functions.

Let us first explore for some cases where closed results are
still available before drawing conclusions for the generic case.

\subsection{$AdS_3\times S^3$ with equal radii}

The kernel for $AdS_3$ can be obtained from the heat kernel for
$H^3$~\cite{Cam90} by analytic continuation, for spacelike intervals both must
coincide. For timelike intervals in $AdS_3$, which are the relevant ones for
 the Penrose limit since the null geodesic  is
always spacelike on the sphere so that it must be timelike in
$AdS_3$~\cite{DS03}, one has the continuation
\begin{equation}
K_{AdS_3}(\zeta\mid s)=\frac{i}{(4\pi
is)^{\frac{3}{2}}}\,\triangle^{\frac{1}{2}}\;e^{i\zeta^2/4s-\,is/a^2},
\end{equation}
where $\frac{\zeta^2}{2}$ is the geodetic interval and
$\triangle^{\frac{1}{2}}=\frac{\zeta/a}{\sinh{(\zeta/a)}} $.
This kernel gives the standard Green
function corresponding to Dirichlet boundary conditions, which can
be expressed in terms of hypergeometric functions (see, e.g.~\cite{DF02}).

This allows us to write the exact kernel for $AdS_3\times S^3$,
given again by a sum to
produce the periodicity on the 3-sphere, with the direct geodesic term
\begin{equation}
K^0_{AdS_3\times S^3}(\zeta,\chi\mid s)=\frac{i}{(4\pi
is)^{3}}\,\frac{\zeta/a}{\sinh{(\zeta/a)}}\,\frac{\chi/a}{\sin{(\chi/a)}}
\;e^{\,i(\zeta^2+\chi^2)/4s}.
\end{equation}

In the Penrose limit, the indirect geodesic contributions are suppressed,
$\zeta^2+\chi^2\rightarrow2\sigma$,
$\frac{\zeta/a}{\sinh{(\zeta/a)}}$ and $\frac{\chi/a}{\sin{(\chi/a)}}$ both
$\rightarrow\frac{u-u'}{\sin(u-u')} $ and
one recovers the plane wave results (\ref{pp_results}) for $d=6$.

\subsection{Conformal coupling}

In $AdS_{p+1}\times S^{q+1}$ with equal radii which is then
conformally flat, for the conformally coupled scalar one gets a
powerlike function in the total chordal distance when mapping to
the massless scalar in flat space. This can also be obtained by a
direct summation of the harmonics on the sphere as shown
in~\cite{DSS03}\footnote{ In fact, in~\cite{DSS03} one also obtain
a powerlike function for a particular mass in the case where the
radii are different, when no conformal map to flat space is
possible. We have managed also to reproduce this result using the
kernels and a nice relation to the conformal situation was found
in terms of a contour integral, see
appendix~\ref{appendixD}.}. The limit agrees with the plane wave
result for the massless case where the Green function is an
inverse power of $\Phi$ (see appendix\ref{appendixA}).

We can accommodate this case in our scheme. Start with $AdS^3\times S^3$ and
use the whole kernel, that is
\begin{equation}
K_{AdS_3\times S^3}(\zeta,\chi\mid s)=\, \sum_{n=-\infty}^{\infty}
K^0_{AdS_3\times S^3}(\zeta,\chi+2\pi n a\mid s).
\end{equation}
For the Weyl invariant scalar, corresponding in this case to $m=0$ one can
take the proper time integral and perform the sum of all direct and indirect
geodesics to get
\begin{equation}
G_{AdS_3\times S^3}(\zeta,\chi\mid
s)\varpropto\frac{1}{[\cos(\chi/a)-\cos(\zeta/a)]^2}\varpropto\frac{1}
{[\mbox{total chordal distance squared}]^2}.
\end{equation}
Now one can take the Penrose limit at any of the two stages, in this final
expression or first in the kernel.

The Weyl coupling case for higher dimension can now be generated
by the ``intertwining'' technique~\cite{Cam90}. Applied to the
kernel one obtains a kernel that produces the desired power in the
total chordal distance for the Green function. Alternative, the
intertwining can be applied directly to the Green function. The
intertwining technique reduces basically to the fact when one can
obtain the kernel or the Green function for the conformally
coupled scalar by just taking derivatives with respect to the
chordal distances. One can start with $AdS_3\times S^1$, taking
partial derivative with respect the chordal distance in $AdS$ one
gets the results for the product space with two dimensions higher
in $AdS$ and taking partial derivative with respect the chordal
distance in the sphere one gets the results for the product space
with two dimensions higher in the sphere\footnote{To cover he
whole range of dimensions for the product space $AdS\times S$
([odd,odd], [odd,even], [even,odd] and [even,even]) one needs in
addition the $S^2$ and $AdS_2$ results, see ~\cite{Cam90}.}. This
way one generates the higher negative powers in the total distance
for the conformally coupled scalar \cite{DSS03}. Again, in the
Penrose limit only the leading term of the direct geodesic
survives the limit.

\section{Conclusions}
Our main result is the explicit construction of the spinor and
vector propagator in the plane wave background (\ref{pp_metric})
arising in a Penrose limit from $AdS_{p+1}\times S^{q+1}$. The
spinor propagator is constructed for generic mass values, the
vector propagator for massless gauge fields in Feynman gauge.

The construction was based on the Schwinger-DeWitt technique. In
general backgrounds via this method one gets only an asymptotic
WKB series with respect to the approach to the light cone. Global
issues for the propagators remain open. For the background under
discussion we could show that the series terminates with its
leading or next-to-leading term. This then strongly suggests that
the resulting expressions are indeed the correct propagators. We
checked this by reproducing the scalar propagator already
constructed in the literature by different methods. In this check
we also explained by the WKB exactness the structural similarity
with the flat space scalar propagator pointed out in \cite{MSS02}.
The propagator in both cases is given by the same function of the
respective geodesic distances up to an additional factor generated
by the nontrivial Van Vleck-Morette determinant of the plane wave
background. This  ordinary determinant for the plane wave can be
shown to be equal to the functional determinant of the quadratic
fluctuations in the path integral formalism~\cite{Ne83}, where
leading-WKB-exactness amounts to the exactness of the Gaussian
approximation for the path integral.\\

Besides the explicit construction in the plane wave geometry, we
made some observations on the relation between both propagators
and kernels to those in spaces from which the plane wave arises in
a Penrose limit. After remarking that the plane wave under study
can also be obtained from ESU, we discussed the limit starting
from known explicit expressions both for the scalars and spinors
in ESU. It turned out that only the leading term in the direct
geodesic contribution survives the limit. This nicely corresponds
with the local nature of the Penrose limit. This picture was
supported by similar observations starting from some special
$AdS\times S$ cases. In addition for the $AdS\times S$ propagators
we were able to explain the distinguished role of a special mass
value for non-Weyl invariant coupling in spaces with different
radii for $AdS_{p+1}$ and $S^{q+1}$ \cite{DSS03}. Just for this
value in the exponent of the Schwinger-DeWitt kernel the term
linear in the proper time cancels and one can explicitly perform the sum.
 A contour integral relates the kernels and propagators for this 
special non-conformally flat (conformal to a spacetime with a conical 
singularity) case to a conformal flat, as shown 
in appendix~\ref{appendixD}.\\

Further study should clarify whether there is a  general theorem
behind. Given a generic plane wave arising in a Penrose limit from
some other spacetime, does then the information on the first few
coefficients of the direct geodesic contribution in the original
spacetime always contain enough information to get the plane wave
propagators? Is the WKB-exactness a generic feature of the 
Penrose limit?\\

Of special interest would also be to find a relation between the  
WKB-exactness of the field theoretic propagators on the plane wave 
(\ref{pp_metric}) and the successful semiclassical description of 
strings in this background~\cite{GKP02, Tse03}.

\acknowledgments
This work was supported in part by the Deutsche Forschungsgemeinschaft(DFG). We 
thank Mario Salizzoni and Christoph Sieg for useful discussions.

\begin{appendix}

\section{Geodesic and chordal distances in $ESU_{n+1}$}
\label{appendixA}

Embedding the $n$-sphere in ($n+1$)-Euclidean space, a point on
the sphere is given by the vector $a\,\hat{\Omega}$, with
\begin{equation}
 \hat{\Omega} = (\cos\alpha \; \cos\beta\,, \;\cos\alpha \;
\sin\beta\,,\; \sin\alpha\,\;
 \hat{\omega})
\end{equation}
where $\hat{\omega}$ is a unit vector on the ($n-2$)-sphere and the
parametrization is as in equation (2.4).
The chordal distance squared $\mu_n(x,x')$ between two points $x$
and $x'$ is related to the arc $\chi_n(x,x')$ (direct geodesic
distance) by
\begin{equation}
1-\frac{\mu_n}{2a^2}=\cos\frac{\chi_n}{a}=
\cos\alpha\,\cos\alpha'\, \cos(\beta-\beta')+\sin\alpha\,
\sin\alpha'\; \cos\frac{\chi_{n-2}}{a}
\end{equation}
where $\cos\frac{\chi_n}{a}\equiv\hat\Omega \cdot \hat\Omega'$ and
$\cos\frac{\chi_{n-2}}{a}\equiv\hat\omega \cdot \hat\omega'$,
being $\chi_{n-2}$ the arc along the ($n-2$)-sphere. Let us take for
simplicity $x'$ to be at the origin.

Going to the local coordinates (equation 2.6) and expanding in inverse
powers of the radius
\begin{eqnarray}
\cos\frac{\chi_n}{a} = \cos\alpha\,\cos\beta=
\cos u-\frac{\Phi}{a^2} + O(a^{-4})
\\
\frac{\chi_n}{a} = u + \frac{\Psi}{a^2} + O(a^{-4}),
\end{eqnarray}
where
\begin{eqnarray}
\Phi=v\sin u + \frac{\vec{x}^2}{2}\,\cos u \\
\Psi=v + \frac{\vec{x}^2}{2}\,\cot u.
\end{eqnarray}

These two quantities naturally arise in the plane wave, $u\Psi$ is the geodetic
interval~\cite{MSS02,KiPi02}
and $2\Phi$ is the limiting value of the total chordal distance squared in
$AdS\times S$ as elucidated in~\cite{DSS03}. Going back to our
$ESU$, it is easy to see that this also holds provided one
compactifies the time into a circle so that $t$ becomes an angle.
That is, as $a\rightarrow \infty$ one has
\begin{eqnarray}
\mbox{geodetic interval}=\,\frac{-a^2t^2+\chi^2}{2}&\rightarrow&
u\Psi \\
\frac{\mbox{total chordal distance squared}}{2}=\, -a^2[1-\cos t]+
\frac{\mu}{2}&\rightarrow&\Phi.
\end{eqnarray}

\section{Spinor Geodesic Parallel Transporter}
\label{appendixB}

Let us go to the frame given by
\begin{eqnarray}
ds^2 &=& 2 \theta^+\theta^- +
\vec{\theta}\cdot\vec{\theta}\equiv\eta_{ab}\theta^a\theta^b\\
\theta^+=du,\qquad\theta^- &=& dv-\frac{1}{2}\vec{x}^2 du,
\qquad\vec{\theta}=d\vec{x}.
\end{eqnarray}
The spin-connection components can be read off from the first Cartan structure
equation
\begin{equation}
d\theta^a + \omega^a_{\,b}\wedge\theta^b
\end{equation}
(tangent indices $a,b=+,-,i$ with $i=1,...,d-2$ being the transverse ones) and
the only nonvanishing ones are
\begin{equation}
\omega^{i-}=-\omega^{-i}=x^i\,du.
\end{equation}

The covariant derivative on spinors
\begin{equation}
\nabla_\mu\equiv\partial_\mu + \Gamma_\mu=\partial_\mu
+\frac{1}{4}\omega_\mu^{ab}\gamma_a\gamma_b\,,
\end{equation}
where the $\gamma 's$ fulfill the Clifford algebra in tangent space
\begin{equation}
 \left\{\gamma_a,\, \gamma_b \right\}=2\eta_{ab}\,\mathbb{I},
\end{equation}
is found to be
\begin{equation}
\nabla_\mu=\left\{\begin{array}{ll}
& \partial_u - \frac{1}{2}\gamma_-\vec{\gamma}\cdot\vec{x}  \\
& \partial_v
\\
& \partial_i\end{array}\right.
\end{equation}
that is, only $\Gamma_u$ is nonzero. An important property is that
$(\Gamma_u)^2=0$ because $(\gamma_-)^2=\mathbb{I}\,\eta_{--}=0$,
i.e. $\Gamma_u$ is nilpotent.

The spinor D'Alembertian can be written in
terms of the scalar one as
\begin{equation}
g^{\mu\nu}\nabla_\mu\nabla_\nu =
\frac{1}{\sqrt{-g}}\partial_\mu(\sqrt{-g}g^{\mu\nu}\partial_\nu) +
2\Gamma_u\partial_v \,.
\end{equation}

The spinor parallel transporter is a bi-spinor that parallel
transports a spinor along a given path and the path we need is the
geodesic connecting the two points. This spinor geodesic parallel transporter
must satisfy the parallel transport equation and the initial condition
\begin{equation}
\partial^\mu\sigma\;\nabla_\mu \mathbb{U}(x,x')=0\,,
\qquad \mathbb{U}(x,x)=\mathbb{I}
\end{equation}
One can write a Dyson-type representation for it (see, e.g.~\cite{Cam92}), 
integrating
along the geodesic emanating from $x'$~\cite{DS03}
 \begin{equation}
\mathbb{U}(t)=\mathbf{P}\exp\,-\int_0^t\Gamma_\mu(\tau) dx^{\mu}(\tau).
\end{equation}
But for the plane wave metric, due to the nilpotency of $\Gamma_\mu$, one can
drop the path ordering symbol $\mathbf{P}$ because the matrices in the
exponent commute, therefore one can perform the integration to get
 \begin{equation}
\mathbb{U}(x,x')=\exp\, \frac{1}{2}\gamma_-\vec{\gamma}\cdot(\vec{x}+\vec{x}'
)\tan\frac{u-u'}{2}
=\mathbb{I}+ \frac{1}{2}\gamma_-\vec{\gamma}\cdot(\vec{x}+\vec{x}'
)\tan\frac{u-u'}{2}.
\end{equation}
Finally, one can easily check that $\square\,\mathbb{U}(x,x')=0$.

\section{Vector Geodesic Parallel Transporter}\label{appendixC}
The Christoffel symbols for the plane wave metric can be directly read off from
the geodesic equations which in turn can be derived from the Lagrangian
 \begin{equation}
\label{lag}
L(\dot{u},\dot{v},\dot{\vec{x}},\vec{x})=\frac{1}{2}\dot{x}^{\mu}\dot{x}_{\mu}
=\dot{u}\dot{v}+\frac{1}{2}
\dot{\vec{x}}^2 -\frac{1}{2}\dot{u}^2\,\vec{x}^2,
\end{equation}
where the dots are derivatives with respect to an affine parameter along the
geodesic. The geodesic equations read
\begin{subequations}
\begin{eqnarray}
\ddot{u}=0\\
\ddot{v}-2\vec{x}\cdot\dot{\vec{x}}\dot{u}=0\\
\ddot{\vec{x}}+\dot{u}^2\vec{x}=0
\end{eqnarray}
\end{subequations}
 and therefore the only nonzero Christoffels  are
 \begin{equation}
\label{Christ}
(\Gamma_u)^i_{\,u}=x^i\,,\qquad (\Gamma_u)^v_{\,i}=(\Gamma_i)^v_{\,u}=-x^i\,.
\end{equation}

There are two types of geodesics~\cite{DS03}: type-A when $\dot{u}=0$ and
the null ones in this category are parallel to the propagation direction of
the wave, and
type-B when one can take $u$ as the affine parameter which is
the generic situation. For this generic case, the Lagrangian (\ref{lag}) is
$\frac{1}{2}\dot{x}^{\mu}\dot{x}_{\mu} = const$  and
reproduces the one for a harmonic oscillator of unit mass and unit 
frequency plus
an extra $\dot{v}$ term. Then, it is not
difficult to see that the recipe to
get the geodetic interval between two generic points is just the
replacing by the classical action for the oscillator between two points
$\vec{x}$ and $\vec{x}'$ followed by the shifts $u\rightarrow u-u'$ and
$v\rightarrow v-v'$, so that

\begin{equation}
\mbox{geodetic interval}=(u-u')(v-v')+(u-u')\left[\frac{\vec{x}^2+
\vec{x}^{'2}}{2}
\,\cot{(u-u')}-\vec{x}\cdot\vec{x'}\,\csc{(u-u')}\right]
\end{equation}
 and for type-A, one just has to let $u\rightarrow 0$ which
 simply produces
\begin{equation}
\mbox{geodetic interval}= \frac{(\vec{x}-\vec{x}')^2}{2}.
\end{equation}
This recipe also works for the quantities $\Psi, \Phi$ and $\triangle$,
previously defined.

The vector parallel transporter is a bi-vector that parallel
transports a vector along a given path, and the path we need is the
geodesic connecting the two points. This vector geodesic parallel transporter
 must satisfy
the parallel transport equation and the initial condition
\begin{equation}
\partial^\rho\sigma\;\nabla_\rho P_{\mu\nu'}(x,x')=0\,,
\qquad P_{\mu\nu}(x,x)=g_{\mu\nu}(x).
\end{equation}
One can also write a Dyson-type representation for it (see, e.g.~\cite{Car97}),
integrating along the geodesic emanating from $x'$~\cite{DS03}
 \begin{equation}
P^{\mu}_{\,\nu'}(x,x')=\mathbf{P}\exp\,-\int_0^t\left(\Gamma_\rho\right)^\mu_
{\,\nu'}(\tau)\, dx^{\rho}(\tau).
\end{equation}
But for the plane wave metric one can check that $\Gamma_\rho$ as a matrix,
with
$\mu$ and $\nu'$ labeling its rows and columns respectively,
commutes with itself at different points. One can
therefore drop the path ordering symbol and perform the integration to get
\begin{subequations}
\begin{eqnarray}
P^{\mu}_{\,\nu'}(x,x')&=&\exp\,\left(\begin{array}{ccc} 0 & 0 &
\vec{0}\\\frac{\vec{x}^2-\vec{x}^{'2}}{2}&0&(\vec{x}+\vec{x}')
\tan\frac{u-u'}{2}
\\-(\vec{x}^{\,t}+\vec{x}'^{\,t})
\tan\frac{u-u'}{2}
&\vec{0}^{\,t}&\mathbb{O}
\end{array}\right)
\\&=&\,\left(\begin{array}{ccc} 1 & 0 &
\vec{0}\\\frac{\vec{x}^2-\vec{x}^{'2}}{2}-\frac{\mid\vec{x}+\vec{x}'\mid^2}{2}
\tan^2\frac{u-u'}{2}&1&(\vec{x}+\vec{x}')
\tan\frac{u-u'}{2}\\
-(\vec{x}+\vec{x}')
\tan\frac{u-u'}{2}&\vec{0}^{\,t}&\mathbb{I}
\end{array}\right).
\end{eqnarray}
\end{subequations}

\section{Non-conformally flat background}
\label{appendixD} Let us consider the Euclidean version $H^3\times
S^3$ with different radii (say, $a$ and $\alpha a$). Up to
normalization factors, the kernel $K^*_{H^3\times
S^3}(\zeta,\chi\mid s)$ is given by
\begin{equation}
\frac{1}{s^3}\,\frac{\zeta/a}{\sinh{(\zeta/a)}}\,\frac{1}{\sin{(\chi/\alpha
a)}} \;e^{\,is/a^2(1/\alpha^2-1)}\,
\sum_{n=-\infty}^{\infty}(\chi/\alpha a \,+\,2\pi
n)\;e^{\,i[\zeta^2+(\chi+2\pi n\alpha a)^2]/4s}.
\end{equation}
The kernel this time has a remaining ``s'' dependent term in the
exponent that can only be eliminated by a special value of the
mass, $m_*^2=\frac{1}{a^2}(1-\frac{1}{\alpha^2})$. This value of
the mass is precisely the one used in ~\cite{DSS03} to get a
closed expression for the Green function. What one can see is
that for this value one can perform the integral to get for the
Green function
\begin{equation}
\frac{\zeta/a}{\sinh{(\zeta/a)}}\,\frac{1}{\sin{(\chi/\alpha a)}}
\; \sum_{n=-\infty}^{\infty}\frac{\chi/\alpha a \,+\, 2\pi
n}{[\zeta^2+(\chi + 2\pi n\alpha a)^2]^2}
\end{equation}
and the resulting series can be exactly computed with the aid of
a Poisson summation (after taking partial derivative with
respect to $x$ to relate both sums)
\begin{equation}
\sum_{n=-\infty}^{\infty}\frac{y}{y^2+(x+n)^2}=\frac{1-e^{-4\pi
y}}{1-2\cos{(2\pi x)}e^{-2\pi y}+e^{-4\pi y}}
\end{equation}
to get
\begin{equation}
G^*_{H^3\times S^3}(\zeta,\chi)\varpropto\frac{\sinh{(\zeta/\alpha
a)}}{\sinh{(\zeta/a)}}\frac{1} {[\cosh(\zeta/\alpha
a)-\cos(\chi/\alpha a)]^2}.
\end{equation}

Now, when the two radii are equal ($\alpha=1$) one gets of course
the conformally coupled scalar in the conformally flat background.
The conformally flat case is periodic on the arc in the sphere
$\chi$ with period $2\pi a$ while the period in the
non-conformally flat is $2\pi \alpha a$. The interesting thing to
notice is that the kernels as well as the Green functions are
related by a contour integral due to Sommerfeld that restores the
appropriate periodicity (see, e.g.~\cite{Do87}). 
This can be explicitly checked for the
special mass above~\cite{MS97}

\begin{equation}
G^*_{H^3\times S^3}(\zeta,\chi)=G_{H^3\times
S^3}(\zeta,\chi)+\frac{i}{4\pi\alpha}\int_\Gamma dw
\,\cot{(\frac{w}{2 \alpha })}\;G_{H^3\times S^3}(\zeta,\chi+ w a),
\end{equation}
where the contour $\Gamma$ consists of two vertical lines from
$(-\pi+i\infty)$ to $(-\pi-i\infty)$ and from $(\pi-i\infty)$ to
$(\pi+i\infty)$ and intersecting the real axis between the poles
of $\cot{(\frac{w}{2 \alpha })}$: $-2\pi\alpha, 0$ and $0,
2\pi\alpha$, respectively.

This very same formula gives the heat kernel for the cone starting
with the one for the plane. This is a remarkable property since
equal radii corresponds to a conformally flat situation and
different radii is conformal to a singular background with a tip,
a conical singularity.
\end{appendix}

\end{document}